\documentclass[twocolumn,amsmath,amssymb,aps,prb,floatfix,showpacs,citeautoscript]{revtex4-1}

\usepackage{graphicx}% Include figure files
\usepackage{dcolumn}% Align table columns on decimal point
\usepackage{bm}% bold math

\def\ket#1{\vert#1\rangle}
\def\bra#1{\langle#1\vert}
\def\ip#1#2{\langle#1\vert#2\rangle}

\def\beq{\begin{equation}}
\def\eeq{\end{equation}}

\def\J{{\bf J}}
\def\P{{\bf P}}
\def\M{{\bf M}}
\def\eb{\bm{\mathcal{E}}}
\def\bb{{\bf B}}
\def\mord{{\cal M}}

\def\Ib{{\bf I}}

\def\q{{\bf q}}
\def\k{{\bf k}}
\def\w{\omega}

\def\chie{\chi^{\rm e}}

\def\chiem{\chi^{\rm em}}
\def\chime{\chi^{\rm me}}
\def\chiq{\chi^{\rm q}}
\def\chiqtil{\wt{{\chi}^{\rm q}}}

\def\sigmas{\sigma^{\mathrm{S}}}
\def\sigmaa{\sigma^{\mathrm{A}}}
\def\sigmah{\sigma^{\mathrm{H}}}

\def\eps{\epsilon}

\def\im{\mathrm{Im}}
\def\re{\mathrm{Re}}

\def\r{{\bf r}}
\def\v{{\bf v}}
\def\k{{\bf k}}

\def\h0{H^0}
\def\A{\mathbf{A}}

\def\wt#1{\widetilde{#1}}

\def\part{\wt{\partial}}

\newcommand{\equ}[1]{Eq.~(\ref{eq:#1})}
\newcommand{\eqs}[2]{Eqs.~(\ref{eq:#1}) and (\ref{eq:#2})}

\begin{document}

\title{Band theory of spatial dispersion in magnetoelectrics}

\author{Andrei Malashevich}
\email{andreim@berkeley.edu}
\author{Ivo Souza}
\affiliation{
Department of Physics, University of California, Berkeley, California 94720, USA
}

\date{\today}

\begin{abstract}
Working in the crystal-momentum representation, we calculate the
optical conductivity of noncentrosymmetric insulating crystals
at first order in the wave vector of light.  The time-even part
of this tensor describes natural optical activity and the
time-odd part describes nonreciprocal effects such as
gyrotropic birefringence. The time-odd part can be uniquely
decomposed into magnetoelectriclike and purely quadrupolar
contributions.  The magnetoelectriclike component reduces in the
static limit to the traceless part of the frozen-ion static 
magnetoelectric polarizability while at finite frequencies it
acquires some quadrupolar character in order to remain
translationally invariant.
The expression for the orbital contribution to the conductivity 
at transparent frequencies
is validated by comparing numerical tight-binding calculations for
finite and periodic samples.
\end{abstract}

% PACS, the Physics and Astronomy Classification Scheme.
% 75.85.+t Magnetoelectric effects, multiferroics
% 42.50.Ct Quantum description of interaction of light and matter;
         % related experiments 
% 42.50.Nn Quantum optical phenomena in absorbing, 
         % amplifying, dispersive and conducting media;
         % cooperative phenomena in quantum optical systems
% 78.20.-e Optical properties of bulk materials and thin films          
% 78.20.Bh Theory, models, and numerical simulation (optics)
% 78.20.Ek Optical activity
% 78.20.Fm Birefringence

\pacs{78.20.Ek,75.85.+t,78.20.Bh}

\maketitle
%%%%%%%%%%%%%%%%%%%%%%%%%%%%%%%%%%%
\marginparwidth 2.7in
\marginparsep 0.5in
\def\amm#1{\marginpar{\small AM: #1}}
\def\ism#1{\marginpar{\small IS: #1}}
\def\scr{\scriptsize}
%%%%%%%%%%%%%%%%%%%%%%%%%%%%%%%%%%%

\section{Introduction}

Electric and magnetic effects are closely coupled in magnetoelectric
(ME) materials. These are insulators with broken spatial-inversion
($\mathcal{P}$) and time-reversal ($\mathcal{T}$) symmetries, in which
an applied electric field $\eb$ induces a first-order magnetization
$\M$, and conversely a magnetic field $\bb$ induces a first-order
electric polarization $\P$.  This cross response is described in the
static limit by a single magnetoelectric polarizability tensor
\beq
\label{eq:ME}
\alpha_{ab}\equiv\frac{\partial M_b}{\partial \mathcal{E}_a}=
\frac{\partial P_a}{\partial B_b},
\eeq
where the equality follows from changing the order of the mixed derivatives
of the free energy.

The ME effect has been intensively studied in recent years.  While the
focus has been mostly on the static response, ME effects in the
optical range have also been observed.\cite{arima08} For oscillating
fields the thermodynamic argument leading to the second equality in
\equ{ME} does not hold because the system is not in equilibrium, and
two separate frequency-dependent polarizabilities are needed to
describe the dynamical ME coupling
\beq
\label{eq:dynME}
\chi^{\mathrm{me}}_{ab}=\frac{\partial M_a}{\partial \mathcal{E}_b},\mbox{~~~}
\chi^{\mathrm{em}}_{ab}=\frac{\partial P_a}{\partial B_b}.
\eeq
It was recognized already in the 1960s that the coupling,
Eq.~(\ref{eq:dynME}), leads to new optical effects in ME media, such as
gyrotropic birefringence.\cite{brown63} Since the lattice-mediated
response is frozen out at optical frequencies, the purely electronic
contribution can be isolated. The first successful measurements, on
Cr$_2$O$_3$, found that the strength of the optical ME coupling is
comparable to that of the static one.\cite{krichevtsov93}

The phenomenology of optical ME effects has been studied in
detail in the literature, starting with the work of Hornreich and
Shtrikman on gyrotropic birefringence.\cite{hornreich68} These
authors showed that this effect is a consequence of spatial
dispersion, appearing at first order in the expansion of the
effective optical conductivity tensor (defined by
Eq.~(\ref{eq:sigma_ab}) below) in powers of the wave vector $\q$ of
light
\beq
\label{eq:sigma-taylor}
\sigma_{ab}(\q,\w)=\sigma^{(0)}_{ab}(\w)
+\sigma_{abc}(\w)q_c
+\cdots
\eeq
It is well known that the phenomenon of natural optical activity is
also a manifestation of spatial dispersion.\cite{landau} While natural
optical activity is associated with the $\mathcal{T}$-even part of
$\sigma_{abc}(\w)$, optical ME effects arise from the
$\mathcal{T}$-odd part, which can be nonzero only in magnetically
ordered systems, where $\mathcal{T}$ symmetry is spontaneously broken.
A careful consideration of all response tensors which contribute to
the conductivity at linear order in $\q$ shows that these include, in
addition to the dynamic ME polarizabilities, Eq.~(\ref{eq:dynME}), the
electric-quadrupole response of the medium.

Regarding the microscopic theories needed for quantitative
calculations, there are well-established {\it molecular} theories of
spatial dispersion,\cite{barron2004,raab2005} but the corresponding
theory for crystals is not equally developed.  A band theory of
natural optical activity was put forth by Natori\cite{natori75} but
has not been used in first-principles calculations.  To our knowledge,
only one group has reported calculations of natural optical activity
in solids at optical wavelengths, based on a
somewhat different formulation.\cite{zhong92,zhong93} As for the optical
ME effects, quantitative estimates of their magnitude have so
far relied on cluster models to mimic the crystalline
environment.\cite{muthkumar-prb96,igarashi-prb09}

In this work, we develop a formalism for calculating
spatial-dispersion effects in the framework of band theory. One
difference with respect to previous works is that
we give a unified treatment of both ${\cal T}$-even and ${\cal
T}$-odd parts of this tensor. More importantly, we express the
transition matrix elements in the crystal momentum
representation.\cite{blount62} This choice has both practical and
formal advantages.  The practical advantage is that it leads to
expressions which can be easily implemented
using localized Wannier orbitals.  
On the theoretical side, the
crystal-momentum representation is the language in which the modern
theories of electric polarization,\cite{King-Smith,resta-review07}
orbital magnetization,\cite{timo05,xiao05,ceresoli06,shi07} and
orbital magnetoelectric response \cite{malash2010,essin2010} are
formulated.
As we shall see, our expression for the orbital contribution to
the $\mathcal{T}$-odd part of $\sigma_{abc}(\w)$ generalizes to finite
frequencies the traceless part of the orbital ME polarizability
formula of Refs.~\onlinecite{malash2010,essin2010}.

The manuscript is organized as follows.  In Sec.~\ref{sec:phenom}
we give a self-contained account of the phenomenology of
spatial-dispersion optics.  The effective conductivity is defined and
related to the magnetoelectric and quadrupolar polarizabilities. We
then reformulate the phenomenological relations, originally obtained
for finite systems, in terms of translationally invariant quantities
which remain well defined in the thermodynamic limit.  The main
results of the paper are contained in Sec.~\ref{sec:bulk}, where we
obtain a microscopic expression for the $\sigma_{abc}(\w)$ in
periodic insulators. We then consider the
$\omega\rightarrow 0$ limit of that expression and discuss its
relation to the theory of static ME response.  In
Sec.~\ref{sec:results} we implement the bulk $\sigma_{abc}(\w)$
expression for a tight-binding model
and compare the results with calculations on finite samples
cut from the bulk crystal.  We conclude in Sec.~\ref{sec:sum} with
a brief summary and outlook.

%----------------------
\section{Phenomenology of spatial dispersion}
\label{sec:phenom}
%----------------------

In this section we discuss spatial dispersion from a phenomenological
perspective.  Besides introducing basic definitions and setting the
notation, the main purpose here is to arrive at
Eqs.~(\ref{eq:alpha_gamma_def})--(\ref{eq:beta-tilde}) relating the
spatially dispersive optical conductivity to 
{\it translationally invariant} renormalized multipole
polarizabilities. Those relations will allow us to identify the
magnetoelectriclike and purely quadrupolar parts of the optical
response of crystals, to be calculated in Sec.~\ref{sec:bulk}.

\subsection{Effective conductivity tensor}

Consider a crystal with broken $\mathcal{P}$ and possibly broken
$\mathcal{T}$ symmetries.  We are mainly interested in materials where
those symmetries are broken spontaneously, rather than by static
electric and magnetic fields, and wish to study their current response
${\bf J}(\q,\w)$ to an electromagnetic plane wave
\beq
\label{eq:waveE}
\eb(\r,t)=\eb(\q,\w)e^{i(\q\cdot\r-\w t)}, 
\eeq
\beq
\label{eq:waveB}
\bb(\r,t) 
=\frac{c}{\w}\big[\q\times\eb(\q,\w)\big]e^{i(\q\cdot\r-\w t)}.
\eeq
Because the oscillating electric and magnetic fields $\eb$ and ${\bm{\mathcal{\epsilon}}}$ and $\bb$
are interdependent, 
the linear (in the field strengths) response can be described by a
single {\it effective conductivity} tensor\cite{hornreich68,melrose}
\beq
\label{eq:sigma_ab}
J_a(\q,\w)=\sigma_{ab}(\q,\w)\mathcal{E}_b(\q,\w).
\eeq
Alternatively, one may choose to work with the dielectric function
$\epsilon_{ab}(\q,\w)$.\cite{landau,melrose} To first order
in $\q$ the two are related (in Gaussian cgs units) by
\beq
\label{eq:epsilon}
\epsilon_{ab}(\q,\w)=\delta_{ab}+\frac{4\pi i}{\omega}\sigma_{ab}(\q,\w).
\eeq

The leading term in the expansion of
$\sigma_{ab}(\q,\w)$ in powers of $\q$, Eq.~(\ref{eq:sigma-taylor}),
is the optical conductivity in
the electric-dipole approximation. We shall focus on the next term in
the expansion, $\sigma_{abc}$, which is chiefly responsible for
spatial dispersion. Because spatial inversion takes $\q$ into $-\q$, the tensor
$\sigma_{abc}(\w)$ necessarily vanishes in centrosymmetric systems.
Its symmetric ($\sigmas_{abc}$) and antisymmetric
($\sigmaa_{abc}$) parts under the interchange of the first two indices
are, respectively, odd and even under
$\mathcal{T}$.\cite{explan-onsager} The $\mathcal{T}$-even piece
describes natural optical activity, and the $\mathcal{T}$-odd
piece describes non-reciprocal optical effects.
These include, in addition to gyrotropic
birefringence, directional dichroism\cite{arima08} and magnetochiral
effects in chiral ferromagnets.\cite{train08}

Unlike the spontaneous magneto-optical effects coming from the
$\mathcal{T}$-odd part of $\sigma_{ab}^{(0)}$ (magnetic circular
dichroism and birefringence), which require ferromagnetic or ferrimagnetic
order, gyrotropic birefringence can also occur in antiferromagnets
such as Cr$_2$O$_3$.
This is a well-known magnetoelectric material, and indeed the
physical basis for spatial dispersion rests in part on the
magnetoelectric effect.

\subsection{Multipole theory for finite systems}

The connection between spatial dispersion and the magnetoelectric
effect can be readily established by expressing $\J(\q,\w)$ in terms of
the multipole moments of the charge and current distributions.  We
begin by taking the spatial Fourier transform of the current density,
\beq
\J(\q,t)=\frac1V\int d\r e^{-i\q\cdot\r} \J(\r,t)
\eeq
and expanding in powers of $\q$, 
\beq
\J(\q,t)=\J^{(0)}(t)+\J^{(1)}(\q,t)+{\cal O}(q^2).
\eeq
Standard multipole-expansion manipulations\cite{melrose} involving the
continuity equation and integrations by parts show that
$J_a^{(0)}(t)=\partial_tP_a(t)$ and
\beq
\label{eq:J1}
J_a^{(1)}(\q,t)=-\frac{iq_b}{2}\partial_tQ_{ab}(t)+i\epsilon_{abc}cq_bM_c(t),
\eeq
where  $\epsilon_{abc}$ is the antisymmetric tensor of rank three and
${\bf P}$, ${\bf Q}$, and ${\bf M}$ are the electric dipole,
electric quadrupole, and magnetic dipole moments of the sample divided
by its volume
\beq
\label{eq:P}
P_a(t)=\frac1V\int d\r\, r_a\rho(t,\r),
\eeq
\beq
\label{eq:Q}
Q_{ab}(t)=\frac1V\int d\r\, r_ar_b\rho(t,\r),
\eeq
\beq
\label{eq:M}
M_a(t)=\frac{1}{2cV}\epsilon_{abc}\int d\r\, r_bJ_c(t,\r).
\eeq
Fourier transforming in time we arrive at
\beq
\label{eq:J_PQM}
J_a(\q,\w)=-i\w P_a(\w)-\frac{\w}{2}q_bQ_{ab}(\w)
+i\epsilon_{abc}cq_bM_c(\w)+{\cal O}(q^2).  
\eeq

The current induced by the monochromatic wave,
Eqs.~(\ref{eq:waveE}) and (\ref{eq:waveB}), can now be calculated from the
oscillating induced moments, which are the real parts of the following
expressions:\cite{barron2004,raab2005}
\beq
\label{eq:inducedP}
P_a=\chie_{ab}\mathcal{E}_b+\frac12\chiq_{abc}\nabla_c\mathcal{E}_b
+\cdots+\chiem_{ab}B_b+\cdots,
\eeq
\beq
\label{eq:inducedQ}
Q_{ab}=\chiqtil_{abc}\mathcal{E}_c+\cdots,
\eeq
\beq
\label{eq:inducedM}
M_a=\chime_{ab}\mathcal{E}_b+\cdots, 
\eeq
where the fields and their gradients are evaluated at the location of
the sample.  $\chie$ is the electric polarizability per unit volume,
and quantum-mechanical expressions for the remaining response
tensors are listed in Appendix~\ref{app:multipol}. $\chiem$ and
$\chime$ are the dynamic ME polarizabilities introduced in
Eq.~(\ref{eq:dynME}); they involve matrix elements of the
electric-dipole ($E1$) and magnetic-dipole ($M1$) operators, and for
this reason are known as the $E1.M1$ terms.  $\chiq$ and $\chiqtil$ are
the $E1.E2$ terms, as they mix electric-dipole and electric-quadrupole
transitions.

In Eqs.~(\ref{eq:inducedP})--(\ref{eq:inducedM}) only those terms
which contribute to the effective conductivity up to first order in $\q$
were kept.  Combining Eqs.~(\ref{eq:J_PQM})--(\ref{eq:inducedM}) with
Eqs.~(\ref{eq:waveE}) and (\ref{eq:waveB}) and comparing with
Eqs.~(\ref{eq:sigma_ab}) and (\ref{eq:sigma-taylor}) we find, upon
collecting terms linear in $\q$,
\beq
\label{eq:sigma-abc}
\sigma_{abc}=
ic(\chiem_{ad}\epsilon_{dbc}+\epsilon_{acd}\chime_{db})
+\frac{\w}{2}(\chiq_{abc}-\chiqtil_{acb})
\eeq
Spatial dispersion is thus governed by the magnetoelectric 
and quadrupolar 
responses of the medium.\cite{hornreich68}
The need to include the quadrupolar terms in order to properly describe
the optical activity of oriented molecules and uniaxial crystals
was emphasized in Ref.~\onlinecite{buckingham71}.

Dividing Eq.~(\ref{eq:sigma-abc}) into symmetric (magnetic) and
antisymmetric (natural) parts under $a\leftrightarrow b$ yields
\beq
\label{eq:sigma-S}
\sigmas_{abc}=ic
\left(
   \eps_{bcd}\alpha_{ad} 
  +\eps_{acd}\alpha_{bd}   
\right)
+
\omega\gamma_{abc},
\eeq
\beq
\label{eq:sigma-A}
\sigmaa_{abc}=ic
\left(
   \eps_{bcd}\beta_{ad} 
  -\eps_{acd}\beta_{bd}  
\right)+
\omega\xi_{abc},
\eeq
where we have defined
\beq
\label{eq:alpha}
\alpha_{ab}=\frac{\chiem_{ab}+\chime_{ba}}{2}\doteq\re\,\chiem_{ab},
\eeq
which reduces to Eq.~(\ref{eq:ME}) in the static limit, and
\beq
\label{eq:beta}
\beta_{ab}=\frac{\chiem_{ab}-\chime_{ba}}{2}\doteq i\im\,\chiem_{ab},
\eeq
\beq
\label{eq:gamma}
\gamma_{abc}
=\frac{\chiq_{abc}+\chiq_{bac}-\chiqtil_{acb}-\chiqtil_{bca}}{4}
\doteq \frac{i}{2}\im\left[\chiq_{abc}+\chiq_{bac}\right], 
\eeq
\beq
\label{eq:xi}
\xi_{abc}
=\frac{\chiq_{abc}-\chiq_{bac}-\chiqtil_{acb}+\chiqtil_{bca}}{4}
\doteq \frac12\re\left[\chiq_{abc}-\chiq_{bac}\right]. 
\eeq
In each of these equations the second equality, denoted by the symbol
$\dot=$, only holds at nonabsorbing frequencies, for which
$\chiem_{ab}\dot=(\chime_{ba})^*$ and
$\chiqtil_{abc}\dot=(\chiq_{cab})^*$ (see
Appendix~\ref{app:multipol}). In this lossless regime $\sigma_{abc}$
becomes anti-Hermitian in the first two indices.

The above multipole formulation leads to a practical scheme for
calculating spatial dispersion effects, by computing the
polarizabilities $\chiem$, $\chime$, $\chiq$, and $\chiqtil$ from
Eqs.~(\ref{eq:chiem-qm-re})--(\ref{eq:LAM-qm-im}), and assembling them
in Eq.~(\ref{eq:sigma-abc}). This approach can be used for molecules
and other finite systems but not for bulk crystals, because the
quantum-mechanical expressions in Appendix~\ref{app:multipol} become
ill-defined under periodic boundary conditions.  

The problem can be traced back to the integrations by parts carried
out around \equ{J1}, where the boundary terms were discarded. Such
procedure is allowed for finite systems, as the boundary can always be
placed outside the sample. It cannot, however, be rigorously justified
for periodic crystals with delocalized electrons.  This is a subtle
but by now well-understood problem. For example, the macroscopic
electric polarization and orbital magnetization of crystals cannot be
calculated under periodic boundary conditions as the first moments of
the charge and orbital current distributions in one crystalline cell
because the result depends on the choice of cell.\cite{resta-review07}
The correct band-theory expressions for ${\bf P}$ and orbital ${\bf
  M}$ have been derived in Ref.~\onlinecite{King-Smith} and
Refs.~\onlinecite{timo05,xiao05,ceresoli06,shi07}, respectively.

\subsection{Translationally invariant polarizabilities}
\label{sec:transl-inv}

Already for finite systems the description based on
Eqs.~(\ref{eq:inducedP})--(\ref{eq:inducedM}) is highly redundant, as
the individual polarizabilities are
origin dependent.\cite{barron2004,raab2005} The combination of
polarizabilities on the right-hand side of Eqs.~(\ref{eq:sigma-S}) and
(\ref{eq:sigma-A}) is of course translationally invariant (the
conductivity is a physical observable)
but we shall go one step further and redefine the polarizability
tensors 
so that they become {\it individually} origin independent, and hence
well defined for periodic crystals.

To begin, we note that the trace of $\alpha$ drops out from
Eq.~(\ref{eq:sigma-S}), leaving eight magnetoelectric
quantities. These fully specify $\sigmas_{abc}$ in the static limit
while at finite frequencies the quadrupolar tensor
$\gamma_{abc}=\gamma_{bac}$ contributes 18 additional quantities.
This brings the total number to 26, while $\sigmas_{abc}$ itself,
being symmetric in the first two indices, only contains 18 independent
quantities.  The source of this discrepancy lies in the
origin-dependence of the tensors $\alpha$ and $\gamma$, and it can be
removed by suitably redefining them. To that end we note that any
third-rank tensor $\sigmas_{abc}$ symmetric under $a\leftrightarrow b$
can be uniquely expanded as
\beq
\label{eq:alpha_gamma_def}
\sigmas_{abc}=ic
\left(
   \eps_{bcd}\wt{\alpha}_{ad} 
  +\eps_{acd}\wt{\alpha}_{bd}   
\right)
+
\omega\wt{\gamma}_{abc},
\eeq
where 
\beq
\label{eq:alpha-tilde}
\begin{split}
\wt{\alpha}_{da}&=\frac{1}{3ic} \sigmas_{dbc}\eps_{bca}\\
&=\alpha_{da}-\frac{1}{3}{\rm Tr}[\alpha]\delta_{ad}
+\frac{\omega}{3ic}\gamma_{dbc}\epsilon_{bca}
\end{split}
\eeq
(here $\delta_{ad}$ is the Kronecker delta) and 
\beq
\label{eq:Gamma-tilde}
\begin{split}
\wt{\gamma}_{abc}&=\frac{1}{3\w}
\left(
\sigmas_{abc}+\sigmas_{cab}+\sigmas_{bca}
\right)\\
&=\frac13
\left(
\gamma_{abc}+\gamma_{cab}+\gamma_{bca}
\right).
\end{split}
\eeq

Replacing Eq.~(\ref{eq:sigma-S}) with Eq.~(\ref{eq:alpha_gamma_def}) removes
the above-mentioned discrepancy, because the totally symmetric tensor
$\wt{\gamma}_{abc}$ has only ten independent quantities, compared to 18
in $\gamma_{abc}$.  As for the tensor $\wt{\alpha}$, it reduces in the
static limit to the traceless part of the magnetoelectric tensor
$\alpha$. But while $\alpha$ becomes origin dependent at finite
frequencies,\cite{raab2005} $\wt{\alpha}$ remains origin independent
by admixing some quadrupolar character.  
It seems appropriate to interpret the renormalized property
tensor $\wt{\alpha}$ as the traceless {\it optical magnetoelectric
  tensor}, and $\wt{\gamma}$ as the purely quadrupolar part of
$\sigmas_{abc}$. 

We now turn briefly to $\sigmaa_{abc}$.  
A third-rank tensor antisymmetric
in two indices has nine independent components, however, there are 18
quantities on the right-hand side of Eq.~(\ref{eq:sigma-A}). We
therefore replace it with
\beq
\label{eq:sigmaa-tilde}
\sigmaa_{abc}=ic
\left(
   \eps_{bcd}\wt{\beta}_{ad} 
  -\eps_{acd}\wt{\beta}_{bd}  
\right),
\eeq
where 
\beq
\label{eq:beta-tilde}
\begin{split}
\wt{\beta}_{ab}&=\frac{1}{4ic}\epsilon_{bcd}
(2\sigmaa_{acd}-\sigmaa_{cda})\\
&=\beta_{ab}+\frac{\omega}{4ic}\epsilon_{bcd}(2\xi_{acd}-\xi_{cda}).
\end{split}
\eeq
Hence natural optical activity, just like gyrotropic birefringence,
is governed by an origin-independent combination of magnetoelectric
($\beta$) and quadrupolar ($\xi$) terms.\cite{buckingham71}
Alternatively, $\wt{\beta}$ can be interpreted as a renormalized
magnetoelectriclike tensor, in the same way as $\wt{\alpha}$.

Equations (\ref{eq:alpha_gamma_def}) and (\ref{eq:sigmaa-tilde}) for
$\sigmas_{abc}$ and $\sigmaa_{abc}$ correspond to Eqs.~(21) and
(30) of Ref.~\onlinecite{hornreich68} while
Eqs.~(\ref{eq:alpha-tilde}), (\ref{eq:Gamma-tilde}), and
(\ref{eq:beta-tilde}) express the translationally invariant 
property tensors $\wt{\alpha}$, $\wt{\beta}$, and $\wt{\gamma}$ as
combinations of origin-dependent multipole polarizabilities.

%-------------------------------------------
\section{Evaluation of the conductivity}
\label{sec:bulk}
%-------------------------------------------

In this section we derive, working in the independent-particle
approximation, a quantum-mechanical expression for $\sigma_{abc}(\w)$.
The expression, valid for band insulators, is conveniently written as
a sum of two terms, which we shall denote by the superscripts (m) and
(e). They arise, respectively, from the $q$ dependence of the transition
matrix elements and of the transition energies.\cite{natori75}

At nonabsorbing frequencies $\sigma_{abc}(\w)$ is an anti-hermitian tensor
in the first two indices.
The imaginary (symmetric) part is given, at $T=0$, by the sum of
\beq
\label{eq:im-sigma-s-delta}
\begin{split}
\im\,\sigma^{({\rm m})}_{{\rm S},abc}(\omega)=\frac{2e^2}{\hbar}\int[d\k]
\sum_{n,l}^{o,e}\,\frac{\w_{ln}}{\w_{ln}^2-\w^2} \\
\times\im\left(A_{ln,b}B_{nl,ac}+A_{ln,a}B_{nl,bc}\right)  
\end{split}
\eeq
and
\beq
\label{eq:im-sigma-s-delta-prime}
\begin{split}
\im\,\sigma^{({\rm e})}_{{\rm S},abc}(\omega)=\frac{2e^2}{\hbar^2}\int[d\k]
\sum_{n,l}^{o,e}\,\frac{\w_{ln}^3}{(\w_{ln}^2-\w^2)^2} \\
\times\partial_c(E_l+E_n)\re\left(A_{nl,a}A_{ln,b}\right),  
\end{split}
\eeq
and the real (antisymmetric) part is the sum of
\beq
\label{eq:re-sigma-a-m}
\begin{split}
\re\,\sigma^{({\rm m})}_{A,abc}(\omega)=\frac{2e^2}{\hbar}\int[d\k]
\sum_{n,l}^{o,e}\,\frac{\w}{\w_{ln}^2-\w^2} \\
\times\re\left(A_{ln,b}B_{nl,ac}-A_{ln,a}B_{nl,bc}\right)  
\end{split}
\eeq
and
\beq
\label{eq:re-sigma-a-e}
\begin{split}
\re\,\sigma^{({\rm e})}_{A,abc}(\omega)=-\frac{e^2}{\hbar^2}\int[d\k]
\sum_{n,l}^{o,e}\,\frac{(3\w_{ln}^2-\w^2)\w}{(\w_{ln}^2-\w^2)^2} \\
\times\partial_c(E_l+E_n)\im\left(A_{nl,a}A_{ln,b}\right).  
\end{split}
\eeq

In these expressions the indices $n$ and $l$ run over occupied ($o$)
and empty ($e$) bands, respectively, $[d\k]$ stands for
$d^3k/(2\pi)^3$, $\partial_c=\partial/\partial_{k_c}$, and
$\hbar\omega_{ln}=E_l-E_n$.  All quantities in the integrands are
labeled by the index $\k$, which has been omitted for brevity. The
matrix $A_{nl,a}=A_{ln,a}^*$, known as the Berry connection, is
defined as
\beq
\label{eq:A}
A_{nl,a}=i\bra{u_n}\partial_a u_l\rangle %=A_{ln,a}^*
\eeq
and the matrix $B_{nl,ac}=-B_{ln,ac}^*$ has both orbital and spin 
contributions, 
\beq
\label{eq:B-ac}
B_{nl,ac}=B_{nl,ac}^{({\rm orb})}+B_{nl,ac}^{({\rm spin})},
\eeq
given by
\beq
\label{eq:B-ac-orb}
B_{nl,ac}^{({\rm orb})}=\frac{1}{2\hbar}
\left[
  \bra{u_n}(\partial_aH)\ket{\partial_c u_l}
 -\bra{\partial_c u_n}(\partial_aH)\ket{u_l}
\right]
\eeq
and
\beq
\label{eq:B-ac-spin}
B_{nl,ac}^{({\rm spin})}=-\frac{i}{m_e}\epsilon_{abc}
\bra{u_n}S_b\ket{u_l},
\eeq
where $u_{n\k}$ is a cell-periodic Bloch state, $H_{\k}$ is
related to the crystal Hamiltonian ${\cal H}$ by $e^{-i\k\cdot\r}{\cal
  H} e^{i\k\cdot\r}$, and  $m_e$ is the electron mass. 

The energy (e) terms have purely orbital character, while the matrix
element (m) terms have both orbital and spin components. It can
  be verified that the spin part of \equ{im-sigma-s-delta} does not
  contribute to \equ{Gamma-tilde}, consistent with the fact that
  $\wt{\gamma}_{abc}$ is a purely orbital (electric-quadrupolar)
  quantity.

\subsection{Derivation}

The derivation of the equations given above proceeds as follows. We
first evaluate the absorptive (Hermitian) part of $\sigma_{abc}$, and then
insert its symmetric and antisymmetric parts into the Kramers-Kr\"onig 
relations
\beq
\label{eq:kk1}
\im\,\sigma_{abc}(\w_0)=-\frac{1}{\pi}\mathrm{P}\int_{-\infty}^\infty\,
\frac{\re\,\sigma_{abc}(\w)}{\w-\w_0}\,d\w
\eeq
and
\beq
\label{eq:kk2}
\re\,\sigma_{abc}(\w_0)=\frac{1}{\pi}\mathrm{P}\int_{-\infty}^\infty\,
\frac{\im\,\sigma_{abc}(\w)}{\w-\w_0}\,d\w,
\eeq
respectively.

The Kubo-Greenwood formula
for the absorptive part of the conductivity 
at finite $\w$ and $\q$ reads
\beq
\label{eq:sigma}
\begin{split}
\sigmah_{ab}&(\q,\w)=\frac{\pi e^2}{\hbar\w}
\int [d\k]\sum_{nl}\,(f_{n,\k-\q/2}-f_{l,\k+\q/2}) \\ 
&\times\bra{\psi_{n,\k-\q/2}}I^\dagger_a(\q)\ket{\psi_{l,\k+\q/2}} 
\bra{\psi_{l,\k+\q/2}}I_b(\q)\ket{\psi_{n,\k-\q/2}} \\
&\times\delta\left[\omega-\omega_{ln\k}(\q)\right],
\end{split}
\eeq
where $f_{n\k\pm \q/2}$ is the occupation factor of the Bloch state
$\psi_{n\k\pm \q/2}$ with eigenenergy $E_{n\k\pm \q/2}$,
\beq
\hbar\omega_{ln\k}(\q)=E_{l,\k+\q/2}-E_{n,\k-\q/2},
\eeq
and ${\bf I}(\q)$ is related to the velocity and spin operators by
\beq
\Ib(\q)=\frac{e^{i\q\cdot\r}\v+\v e^{i\q\cdot\r}}{2}+
\frac{i}{m_{\mathrm{e}}}({\bf S}\times\q)e^{i\q\cdot\r}.
\eeq
Equation~(\ref{eq:sigma}) 
reduces in the limit $\q\rightarrow 0$
to the familiar expression for the optical conductivity in the
electric-dipole approximation.\cite{harrison80}
It can be derived starting from the interaction
Hamiltonian 
\beq
H_{\rm I}=\frac{e}{2c}(\A\cdot\v+\v\cdot\A)+
\frac{e}{m_ec}({\boldsymbol\nabla}\times{\bf A})\cdot{\bf S}.
\eeq 

Up to terms linear in $\q$, the optical matrix element
$\bra{\psi_{n,\k-\q/2}}I_a^{\dagger}(\q)\ket{\psi_{l,\k+\q/2}}$ may be replaced by
\beq
\label{eq:B-q}
\begin{split}
B_{nl\k,a}(\q)&\equiv\bra{u_{n,\k-\q/2}}
v_a(\k)
-\frac{i}{m_e}(S\times q)_a
\ket{u_{l,\k+\q/2}}\\
&=B_{nl,a}^{(0)}+B_{nl,ac}q_c+\cdots
\end{split}
\eeq
where $\v(\k)=e^{-i\k\cdot\r}\v e^{i\k\cdot\r}$.
Using the relation\cite{blount62} $\hbar\v(\k) =\partial_a H_{\k}$ 
together with \equ{A},
the expansion coefficients in the second line are found to be
\beq
\label{eq:B-a}
B_{nl,a}^{(0)}=\bra{u_n}v_a\ket{u_l}
=i\w_{nl}A_{nl,a}+\frac1{\hbar}\delta_{ln}\partial_aE_l 
\eeq
and Eqs.~(\ref{eq:B-ac})--(\ref{eq:B-ac-spin}) for $B_{nl,ac}$.

We are now ready to calculate $\sigmah_{abc}$ by differentiating
\equ{sigma} with respect to $q_c$. Because we assume an insulator at
$T=0$,\cite{explan-opt-act-metals} the derivative acts only on the
transition matrix elements and on the $\delta$ function selecting the
transition energies, not on the occupation factors.
Using \equ{B-q} for the matrix elements [note that the second,
intraband, term in \equ{B-a} does not contribute in insulators],
together with
\begin{equation}
\begin{split}
&\left.
\frac{\partial}{\partial_{q_c}}\delta\left[\omega-\omega_{ln\k}(\q)\right]
\right|_{\q=0}= \\
&\;\;\;\;\;\;\;\;
-\frac{1}{2\hbar}\delta'\big(\omega-\omega_{ln\k}(0)\big)
  \partial_c\left(E_{l\k}+E_{n\k}\right)
\end{split}
\end{equation}
and inserting the result for the symmetric and antisymmetric parts of
$\sigmah_{abc}$ into \equ{kk1} and \equ{kk2} respectively, one easily
obtains Eqs.~(\ref{eq:im-sigma-s-delta})--(\ref{eq:re-sigma-a-e}).

\subsection{Static limit}
\label{sec:static}

In the limit $\w\rightarrow 0$ the ME tensors $\chiem_{ab}$ and
$\chime_{ba}$ become identical, and as a result $\sigmaa_{abc}$
[Eq.~(\ref{eq:sigma-A})] vanishes.  As for $\sigmas_{abc}$, we noted
in Sec.~\ref{sec:transl-inv} that its dc limit is governed by
$\wt{\alpha}(0)$, the traceless part of the static ME polarizability
tensor $\alpha(0)$. Since our calculation of $\sigmas_{abc}$ only
included the purely electronic response to the optical fields, we
should recover in that limit the frozen-ion part of
$\wt{\alpha}(0)$.

We will focus here on the orbital contribution to $\sigmas_{abc}$, and
compare it with the band-theory expression obtained in
Refs.~\onlinecite{malash2010,essin2010} for the frozen-ion orbital ME
tensor.  The corresponding proof for the spin contribution is
elementary.

We begin by recasting the orbital part of
Eqs.~(\ref{eq:im-sigma-s-delta}) and (\ref{eq:im-sigma-s-delta-prime}) at
$\w=0$ in a form where empty states do not appear explicitly. This is
done in Appendix~\ref{app:static}, where we obtain
\beq
\label{eq:sigmaSfin}
\begin{split}
&\im\,\sigma^{\rm (orb)}_{{\rm S},abc}(0)
=\frac{e^2}{\hbar}\int[d\k]\sum_{nm}^o
\re\Bigr\{\ip{\partial_au_n}{\partial_cu_m}\ip{u_m}{\partial_bu_n}\\
&\;\;\;\;\;\;\;\;\;\;\;\;\;\;\;\;\;\;\;\;\;\;\;\;\;\;\;\;\;\;\;\;\;\;\;\;\;\;\;\;\;\;
+\ip{\partial_bu_n}{\partial_cu_m}\ip{u_m}{\partial_au_n}\Bigl\}
\\
&+\frac{e}{\hbar}\int[d\k]\sum_n^o
\Big\{
\left[
\im\bra{\partial_cu_n}\partial_a(H+E_n)\ket{\wt{\partial}_{\mathcal{E}_b}u_n}
- a\leftrightarrow c
\right]\\
&\;\;\;\;\;\;\;\;\;\;\;\;\;\;\;\;\;\;\;\;\;\;\,\;\;
+b\leftrightarrow a
\Big\},
\end{split}
\eeq
where the covariant field derivative
$\ket{\wt{\partial}_{\mathcal{E}_b}u_n}$ is given by \equ{stern2}.
Equation~(\ref{eq:sigmaSfin}) can now be compared with Eq.~(C.2) of
Ref.~\onlinecite{malash2010} for the static ME tensor, which reads
\begin{equation}
\label{eq:al_da}
\begin{split}
\alpha_{da}^{({\rm orb})}&(0)=
\frac{e^2}{2\hbar c}\epsilon_{abc}\int[d\k]\sum_{nm}^o\re\Bigl\{\ip{\partial_bu_n}{\partial_cu_m}\ip{u_m}{\partial_du_n}\Bigr\}\\
&-\frac{e}{\hbar c}\epsilon_{abc}\int[d\k]\sum_n^o
\im\bra{\partial_bu_n}\partial_c(H+E_n)\ket{\wt{\partial}_{\mathcal{E}_d}u_n}.
\end{split}
\end{equation}
It is easily verified that inserting Eq.~(\ref{eq:al_da}) into
Eq.~(\ref{eq:sigma-S}) at $\w=0$ yields Eq.~(\ref{eq:sigmaSfin}),
which proves the result.

%--------------------------
\section{Numerical results}
\label{sec:results}
%--------------------------

\begin{figure}
\centering\includegraphics{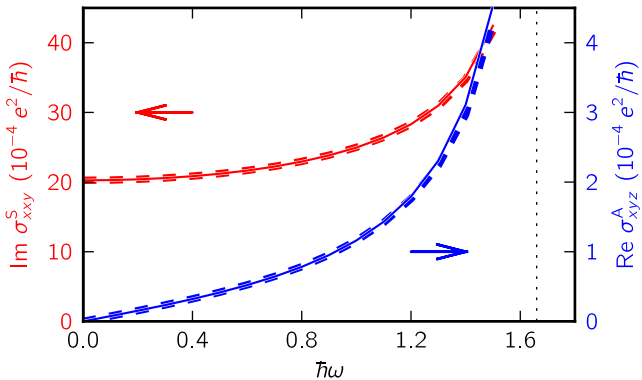}
\caption{(Color online)  The $xxy$ component of the gyrotropic
birefringence tensor 
$\im\,\sigmas_{abc}$, and the $xyz$ component of 
the natural optical activity tensor $\re\,\sigmaa_{abc}$, 
calculated for the tight-binding
model described in the text as a function of frequency.
Solid lines: extrapolation from calculations on finite crystallites.
Dashed lines: calculations on periodic crystals using the $k$-space
formulas derived in this work.  The vertical dotted line indicates the
frequency corresponding to the direct band gap.}
\label{fig:spectrum}
\end{figure}

In order to check the expressions derived in the previous section, we
have carried out numerical tests comparing calculations done under
periodic boundary conditions against reference calculations on finite
crystallites.
We chose for our tests the tight-binding model of
Ref.~\onlinecite{malash2010}. This is a spinless model on a
$2\times2\times2$ cubic lattice, where $\mathcal{P}$ symmetry is
broken by assigning random on-site energies and $\mathcal{T}$ symmetry
is broken by complex first-neighbor hoppings. The model parameters in
Table~A.1 of Ref.~\onlinecite{malash2010} were used (one of the
complex hopping phases, labeled $\varphi$ therein, shall be used as a
control parameter), and the two lowest bands were treated as occupied.

The tensor components
$\im\,\sigmas_{xxy}$ and $\re\,\sigmaa_{xyz}$
were evaluated at
nonabsorbing frequencies.  The calculations on periodic samples were
done on a $30\times 30\times 30$ mesh of $k$ points using
Eqs.~(\ref{eq:im-sigma-s-delta})--(\ref{eq:B-ac-orb}), together with the
sum-over-states formula for ${\boldsymbol\nabla}_\k
\ket{u_{n\k}}$.\cite{malash2010} For the calculations on finite
samples we used \eqs{sigma-S}{sigma-A},
\beq 
\label{eq:sigmas_xxy}
\im\,\sigmas_{xxy}=2c\re\,\alpha_{xz}+\w\im\,\gamma_{xxy} \doteq
2c\alpha_{xz}-i\w\gamma_{xxy}
\eeq 
and
\beq
\begin{split}
\label{eq:sigma-a-xyz}
\re\,\sigma_{xyz}^{\mathrm{A}}&=
-c\im(\beta_{xx}+\beta_{yy})+\w\re\,\xi_{xyz} \\
&\doteq ic(\beta_{xx}+\beta_{yy})+\w\xi_{xyz},
\end{split}
\eeq
together with Eqs.~(\ref{eq:alpha})--(\ref{eq:xi}) and
(\ref{eq:chiem-qm-re})--(\ref{eq:LAM-qm-im}) for the magnetoelectric
($\alpha$, $\beta$) and quadrupolar ($\gamma$, $\xi$) tensors. We
chose cubic samples containing $L\times L\times L$ unit cells, with
$L=1,\, 2,\, 3,\, 4$, and then extrapolated the calculated values 
to $L\rightarrow
\infty$.\cite{malash2010}

Figure~\ref{fig:spectrum} shows as solid (dashed) lines the frequency
dependence of $\im\,\sigmas_{xxy}$ and $\re\,\sigmaa_{xyz}$ for finite
(periodic) samples, with the parameter $\varphi$ set to $\pi$.  The
natural optical activity spectrum
starts off at zero and increases with frequency,
exhibiting a resonant behavior as the minimum direct gap, denoted by
the vertical dashed line, is approached.  The ME optical spectrum
displays a similar behavior, except that it
remains finite as $\w$ goes to zero.  The excellent agreement between
solid and dashed lines demonstrates the correctness of the $k$-space
formulas.

\begin{figure}
\centering\includegraphics{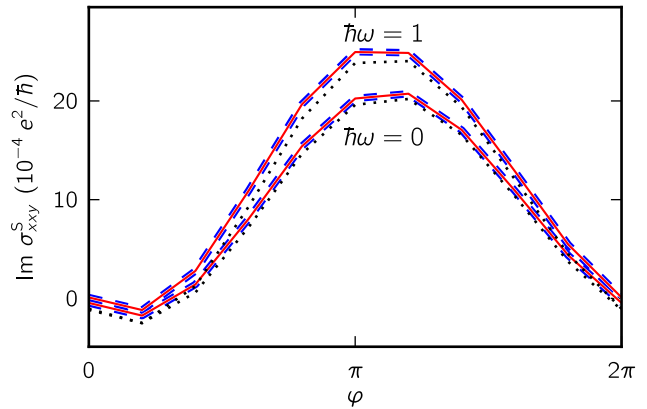}
\caption{(Color online)  The $xxy$ component of 
  $\im\sigmas_{abc}(\w)$, calculated for the tight-binding model
  described in the text as a function of the parameter
  $\varphi$. Solid lines: extrapolation from calculations on finite
  crystallites using \equ{sigmas_xxy}.  Dashed lines: calculations on
  periodic crystals using 
  \eqs{im-sigma-s-delta}{im-sigma-s-delta-prime}.  Dotted lines: same
  as the dashed lines, but ommiting the contribution coming from
  \equ{im-sigma-s-delta-prime}.}
\label{fig:sig_xxy}
\end{figure}

Next we discuss a number of additional numerical tests where we investigate
in more detail the behavior of
$\im\,\sigmas_{xxy}$. In these tests the frequency was kept fixed, and
the parameter $\varphi$ was scanned over the
range $[0,2\pi]$.

In Figure~\ref{fig:sig_xxy} we plot $\im\,\sigmas_{xxy}$ versus
$\varphi$ for two frequencies, $\w=0$ and $\hbar\w=1$.  As before,
solid and dashed lines represent calculations on finite and periodic
samples respectively.  In addition, we show as dotted lines the result
of a periodic-sample calculation using only the matrix element (m)
term, Eq.~(\ref{eq:im-sigma-s-delta}), i.e., omitting the energy (e) term,
\equ{im-sigma-s-delta-prime}.  We see that the energy term gives a
small but visible contribution, which must be included in order
to find agreement with the finite-sample calculation.

\begin{figure}
\centering\includegraphics{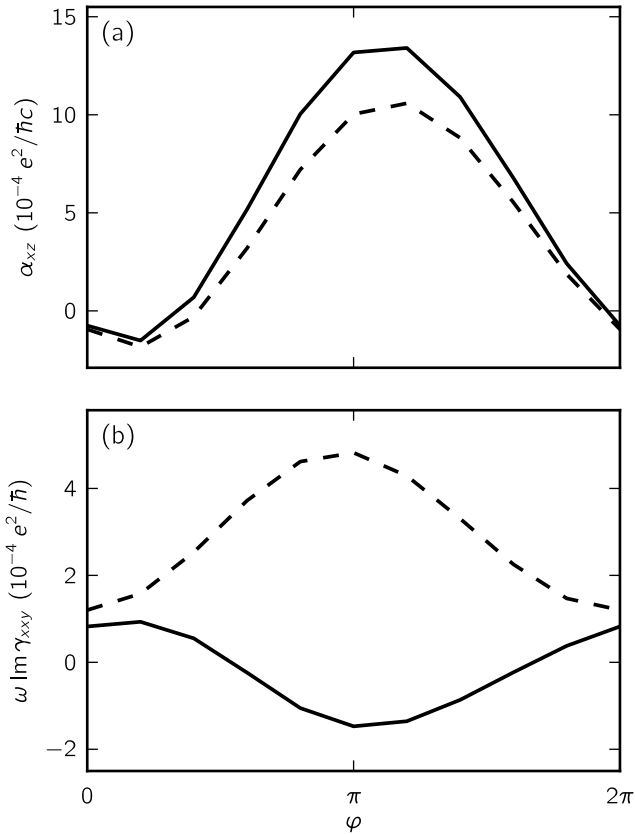}
\caption{Origin-dependence of the bare magnetoelectric (upper
    panel) and quadrupolar (lower panel) polarizabilities appearing on
    the right-hand side of \equ{sigmas_xxy}, calculated at $\hbar\w=1$
    for a finite sample ($L=4$) of the model used in
    Fig.~\ref{fig:sig_xxy}. Solid lines: the center of the sample is
    placed at the origin. Dashed lines: the sample is displaced by
    $\mathbf{r}=(1,1,1)$, in units of the lattice constant
of the $2\times 2\times 2$ cubic cell.}
\label{fig:orig_dep}
\end{figure}

We now turn to the decomposition of  $\im\,\sigmas_{xxy}$ 
according to \equ{sigmas_xxy}, into
magnetoelectric and quadrupolar parts.
They are plotted separately in Fig.~\ref{fig:orig_dep} for $\hbar\w=1$
and $L=4$.  We chose a specific $L$ because $\alpha$ and $\w\gamma$
are origin-dependent quantities, and it is therefore not meaningful to
extrapolate them separately to $L\rightarrow\infty$.  The dashed lines
show how each of them changes when the position of the sample is
shifted. The change in $\alpha_{zz}$ is exactly compensated by the
change in $\w\gamma_{xxy}$, so that the resulting $\im\,\sigmas_{xxy}$
remains the same to machine precision, demonstrating its translational
invariance.

\begin{figure}
\centering\includegraphics{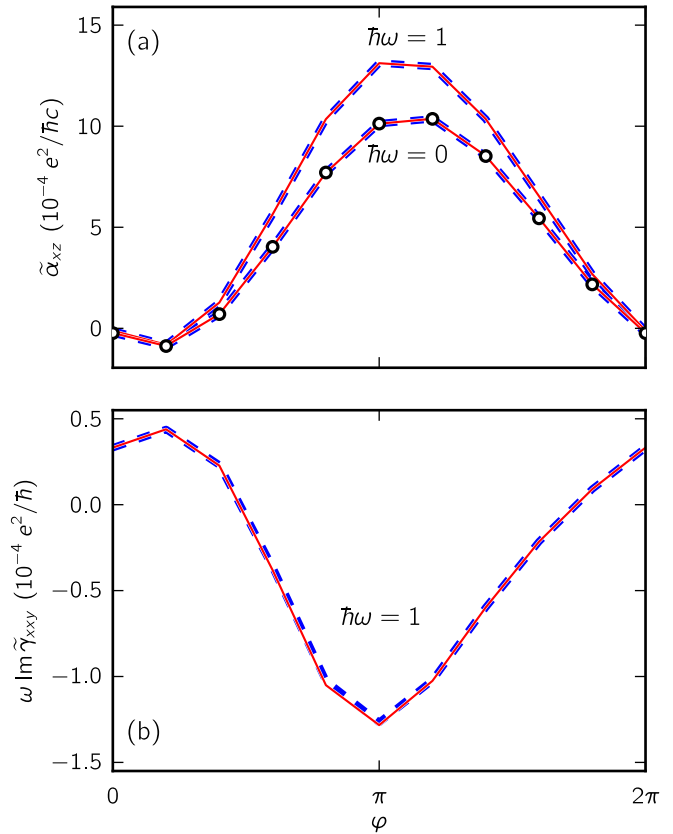}
\caption{(Color online)  Translationally invariant decomposition
  [Eq.~(\ref{eq:sigmas_xxy-inv})] of the curves in
  Fig.~\ref{fig:sig_xxy} into magnetoelectriclike 
  (upper panel) and purely quadrupolar (lower
  panel) contributions.  Solid lines: extrapolation from calculations
  on finite crystallites. Dashed lines: $k$-space calculations
  on periodic crystals. In the static limit the tensor $\wt{\alpha}$
  reduces to the traceless part of the magnetoelectric
  polarizability $\alpha$, and the open circles show
  $\alpha_{xz}(0)$ calculated in $k$ space according to
  Refs.~\onlinecite{malash2010,essin2010}.}
\label{fig:orig_indep}
\end{figure}

An alternative decomposition of $\im\,\sigmas_{xxy}$ is given
by \equ{alpha_gamma_def}:
\beq 
\label{eq:sigmas_xxy-inv}
\im\,\sigmas_{xxy}\doteq
2c\wt{\alpha}_{xz}-i\w\wt{\gamma}_{xxy}.
\eeq 
Unlike the bare property tensors $\alpha$ and $\w\gamma$ appearing in
\equ{sigmas_xxy}, the renormalized magnetoelectriclike and
purely quadrupolar tensors $\wt{\alpha}$ and $\w\wt{\gamma}$ are
origin independent and hence separately well defined for periodic
samples.  Figure~\ref{fig:orig_indep} shows as dashed (solid) lines
their values calculated for periodic (finite) samples from the first
(second) equality in Eqs.~(\ref{eq:alpha-tilde}) and
(\ref{eq:Gamma-tilde}).  Because $\wt{\alpha}$ reduces to the
traceless part of $\alpha$ as $\w\rightarrow 0$, we can directly
compare the curve for $\wt{\alpha}_{xz}(0)$ with a $k$-space
calculation of $\alpha_{xz}(0)$ using the formula derived in
Refs.~\onlinecite{malash2010,essin2010} (open circles).  The precise
agreement confirms numerically the analysis of Sec.~\ref{sec:static}.

%----------------------------
\section{Summary and outlook}
\label{sec:sum}
%----------------------------

In this work we investigated spatial-dispersion optical effects in
insulators. The main result is a band-theory expression for
$\sigma_{abc}(\w)$, the spatially dispersive optical conductivity.
Special attention was given to the ${\cal T}$-odd part of this tensor,
which is nonzero in magnetoelectric crystals, and comprises
magnetoelectriclike ($\wt{\alpha}_{ab}$) and purely quadrupolar
($\wt{\gamma}_{abc}$) contributions.  We showed that each of them
consists of a translationally invariant combination of separately
origin dependent molecular polarizability tensors.

The magnetoelectriclike tensor $\wt{\alpha}_{ab}$ 
 has both spin and
orbital contributions, and the expression for the orbital part
generalizes to finite frequencies the
recently developed band theory of orbital magnetoelectric
response.\cite{malash2010,essin2010} The generalization is, however,
not complete, as the tensor $\wt{\alpha}_{ab}(\w)$ is traceless, and
therefore does not include the isotropic ME coupling.  The reason why
the latter is not recovered from the present formalism is that
our starting point is the current response of an infinite medium to an
electromagnetic wave while the trace of the ME tensor, known as the
{\it axion} contribution, only affects electrodynamics at
boundaries.\cite{hornreich68,essin2010} The calculation of the axion
piece at finite frequencies remains an open problem.

The bulk expression for $\sigma_{abc}(\w)$ at transparent frequencies
was validated by performing numerical calculations on a tight-binding
model, and comparing against reference calculations done on finite
samples.
The quantities needed to evaluate that expression are the occupied and
empty energy eigenvalues and their $k$-space gradients, the
off-diagonal Berry connection matrix Eq.~(\ref{eq:A}), and the orbital
and spin matrices Eqs.~(\ref{eq:B-ac-orb}) and (\ref{eq:B-ac-spin}). The
evaluation of all these objects in a first-principles context can be
done efficiently by mapping the electronic structure onto localized
Wannier orbitals, and then using the technique of Wannier
interpolation.\cite{wang06} This approach has already been used to
compute the magnetic circular dichroism spectrum of
ferromagnets.\cite{yates07}

First-principles calculations of the optical spectrum of solids beyond
the electric-dipole approximation are still in their infancy. We hope
that the formalism introduced in this work will be useful for carrying
out realistic calculations of spatial-dispersion phenomena in the
optical range, including natural optical activity, gyrotropic
birefringence, and directional dichroism.

\acknowledgments
This work was supported by NSF under Grant No. DMR-0706493. Computational 
resources were provided by NERSC.
\appendix

\section{Quantum-mechanical expressions for the 
polarizability tensors}
\label{app:multipol}

In this appendix we list the quantum-mechanical expressions for the
frequency-dependent polarizability tensors $\chiem$, $\chime$, $\chiq$,
and $\chiqtil$ of bounded samples.
They have been used to produce the reference results (solid
  lines) in Figs.~\ref{fig:sig_xxy}--\ref{fig:orig_indep}.

We provide the single-particle version of the formulas in the lossless
regime, which is the form used in Sec.~\ref{sec:results}. A many-body
derivation can be found in Ref.~\onlinecite{raab2005}, and the
modifications needed to describe absorption are discussed in
Refs.~\onlinecite{barron2004,raab2005}.

Defining $Z_{ln}=(V\hbar/2 e^2)(\w_{ln}^2-\w^2)$, where $V$ is the
system volume, the orbital contribution to the magnetoelectric tensor
reads
\beq
\label{eq:chiem-qm-re}
\re\,\chi^{\mathrm{em}}_{ab}\doteq 
\frac{1}{2c}
\sum_{n,l}^{o,e}
\frac{\w_{ln}}{Z_{ln}}
\re\,\bigl[\bra{n}r_a\ket{l}\bra{l}(\r\times\v)_b\ket{n}\bigr]
\doteq\re\,\chi^{\mathrm{me}}_{ba},
\eeq
\beq
\label{eq:chiem-qm-im}
\im\,\chi^{\mathrm{em}}_{ab}\doteq 
\frac{1}{2c}
\sum_{n,l}^{o,e}
\frac{\w}{Z_{ln}}
\im\,\bigl[\bra{n}r_a\ket{l}\bra{l}(\r\times\v)_b\ket{n}\bigr]
\doteq -\im\,\chi^{\mathrm{me}}_{ba}
\eeq
and the quadrupolar polarizability reads
\beq
\label{eq:LAM-qm-re}
\re\,\chiq_{abc}\doteq 
\sum_{n,l}^{o,e}
\frac{\w_{ln}}{Z_{ln}}
\re\,\bigl[\bra{n}r_a\ket{l}\bra{l}r_br_c\ket{n}\bigr]
\doteq\re\,\chiqtil_{cba},
\eeq
\beq
\label{eq:LAM-qm-im}
\im\,\chiq_{abc}\doteq 
\sum_{n,l}^{o,e}
\frac{\w}{Z_{ln}}
\im\,\bigl[\bra{n}r_a\ket{l}\bra{l}r_br_c\ket{n}\bigr]
\doteq -\im\,\chiqtil_{cba}.
\eeq
%

%-----------------------------------------------
\section{Derivation of Eq.~(\ref{eq:sigmaSfin})}
\label{app:static}
%-----------------------------------------------

In order to derive \equ{sigmaSfin}, we drop the spin contribution,
Eq.~(\ref{eq:B-ac-spin}), from
Eqs.~(\ref{eq:im-sigma-s-delta}) and (\ref{eq:im-sigma-s-delta-prime}) and
rewrite the orbital contribution at $\w=0$ as
\begin{equation}
\label{eq:sigmaS}
\im\,\sigma^{({\rm orb})}_{{\rm S},abc}(0)=\frac{e^2}{\hbar}\int[d\k]
\left[
  (C_{abc}+D_{abc})+(C_{bac}+D_{bac})
\right],
\end{equation}
where
\beq
\label{eq:A-abc-static}
\begin{split}
C_{abc}=\sum_{n,l}^{o,e}
\re\biggl\{
\frac{\ip{u_l}{\partial_bu_n}}{E_l-E_n}
\Bigl[
\bra{u_n}(\partial_aH)\ket{\partial_cu_l}-\\
\bra{\partial_cu_n}(\partial_aH)\ket{u_l}
\Bigr]\biggr\}
\end{split}
\eeq
and
\beq
\label{eq:B-abc-static}
D_{abc}=\sum_{n,l}^{o,e}
\frac{\partial_c(E_l+E_n)}{E_l-E_n}
\re\{\ip{\partial_au_n}{u_l}\ip{u_l}{\partial_bu_n}\}.
\eeq

We will use repeatedly the identity\cite{malash2010}
\begin{equation}
\label{eq:stern1}
\partial_c(H-E_l)\ket{u_l}=(E_l-H)\ket{\partial_cu_l},
\end{equation}
as well as the following expression for the field derivative of a
valence-band state projected onto the conduction bands\cite{malash2010}
\begin{equation}
\label{eq:stern2}
\ket{\wt{\partial}_{\mathcal{E}_b}u_n}=
-ie\sum_{l}^e\frac{\ket{u_l}\bra{u_l}}{E_l-E_n}\ket{\partial_bu_n}.
\end{equation}

We start by using Eq.~(\ref{eq:stern1})
to eliminate $\partial_cE_l$ from Eq.~(\ref{eq:B-abc-static}),
\begin{equation}
\begin{split}
D_{abc}&=\sum_{n,l}^{o,e}\re
\biggl\{
\frac{\bra{\partial_au_n}(\partial_cH)\ket{u_l}\ip{u_l}{\partial_bu_n}}{E_l-E_n}\\
&+\frac{\bra{\partial_au_n}H-E_l\ket{\partial_cu_l}\ip{u_l}{\partial_bu_n}}{E_l-E_n}\\
&+\frac{\partial_cE_n}{E_l-E_n}
\ip{\partial_au_n}{u_l}\ip{u_l}{\partial_bu_n}\biggr\}
\end{split}
\end{equation}
and then use Eq.~(\ref{eq:stern2}) twice to find
\begin{equation}
\begin{split}
D_{abc}=-\frac{1}{e}\sum_n^o\im\bra{\partial_au_n}
\partial_c(H+E_n)\ket{\wt{\partial}_{\mathcal{E}_b}u_n}\\
+\sum_{n,l}^{o,e}\re\left\{
  \frac{\bra{\partial_au_n}H-E_l\ket{\partial_cu_l}\ip{u_l}{\partial_bu_n}}{E_l-E_n}\right\}.
\end{split}
\end{equation}
Now write $H-E_l$ as $(H-E_n)+(E_n-E_l)$ and use Eq.~(\ref{eq:stern1}),
\begin{equation}
\label{eq:B}
\begin{split}
D_{abc}&=T_{abc}-\frac{1}{e}\sum_n^o\im\bra{\partial_au_n}
\partial_c(H+E_n)\ket{\wt{\partial}_{\mathcal{E}_b}u_n}\\
&+\sum_{n,l}^{o,e}\re\left\{
\frac{\bra{u_n}\partial_a(E_n-H)\ket{\partial_cu_l}\ip{u_l}{\partial_bu_n}}{E_l-E_n}\right\},
\end{split}
\end{equation}
where we defined
\begin{equation}
\label{eq:Tabc}
T_{abc}=-\sum_{n,l}^{o,e}\re\ip{\partial_au_n}{\partial_cu_l}\ip{u_l}{\partial_bu_n}.
\end{equation}
One term in Eq.~(\ref{eq:B}) exactly cancels the first term in
Eq.~(\ref{eq:A-abc-static}).  For the remainder we use
$\ip{u_n}{\partial_cu_l}=-\ip{\partial_cu_n}{u_l}$ once and then
Eq.~(\ref{eq:stern2}) twice, yielding
\begin{equation}
\begin{split}
&C_{abc}+D_{abc}=T_{abc}\\
&+\frac{1}{e}\sum_n^o\im\left\{
\bra{\partial_cu_n}\partial_a(H+E_n)\ket{\wt{\partial}_{\mathcal{E}_b}u_n}
- a\leftrightarrow c\right\}.
\end{split}
\end{equation}
In order to eliminate the sum over empty states in $T_{abc}$ we need to combine
$C_{abc}+D_{abc}$ with $C_{bac}+D_{bac}$, as in
Eq.~(\ref{eq:sigmaS}). We therefore consider
\begin{equation}
\begin{split}
T_{abc}+T_{bac}&=-\sum_{n,l}^{o,e}\re\Bigl\{
\ip{\partial_au_n}{\partial_cu_l}\ip{u_l}{\partial_bu_n}
+ a\leftrightarrow b
\Bigr\}\\
&=-\sum_n^o\re\Bigr\{\bra{\partial_au_n}(\partial_cQ)\ket{\partial_bu_n}\Bigl\}\\
&=\sum_{nm}^o\re\Bigr\{\ip{\partial_au_n}{\partial_cu_m}\ip{u_m}{\partial_bu_n}\\
&\;\;\;\;\;\;\;\;\;\;\;\;\;
+\ip{\partial_bu_n}{\partial_cu_m}\ip{u_m}{\partial_au_n}\Bigl\},
\end{split}
\end{equation}
where $Q=\sum_l^e\ket{u_l}\bra{u_l}=1-\sum_m^o\ket{u_m}\bra{u_m}$.
Collecting terms, we arrive at Eq.~(\ref{eq:sigmaSfin}).

%\bibliography{pap.bib}

\begin{thebibliography}{10}%
\makeatletter
\providecommand \@ifxundefined [1]{%
 \ifx #1\undefined \expandafter \@firstoftwo
 \else \expandafter \@secondoftwo
\fi
}%
\providecommand \@ifnum [1]{%
 \ifnum #1\expandafter \@firstoftwo
 \else \expandafter \@secondoftwo
\fi
}%
\providecommand \enquote [1]{``#1''}%
\providecommand \bibnamefont  [1]{#1}%
\providecommand \bibfnamefont [1]{#1}%
\providecommand \citenamefont [1]{#1}%
\providecommand\href[0]{\@sanitize\@href}%
\providecommand\@href[1]{\endgroup\@@startlink{#1}\endgroup\@@href}%
\providecommand\@@href[1]{#1\@@endlink}%
\providecommand \@sanitize [0]{\begingroup\catcode`\&12\catcode`\#12\relax}%
\@ifxundefined \pdfoutput {\@firstoftwo}{%
 \@ifnum{\z@=\pdfoutput}{\@firstoftwo}{\@secondoftwo}%
}{%
 \providecommand\@@startlink[1]{\leavevmode}%
 \providecommand\@@endlink[0]{}%
}{%
 \providecommand\@@startlink[1]{%
  \leavevmode
  \pdfstartlink
   attr{/Border[0 0 1 ]/H/I/C[0 1 1]}%
   user{/Subtype/Link/A<</Type/Action/S/URI/URI(#1)>>}%
  \relax
 }%
 \providecommand\@@endlink[0]{\pdfendlink}%
}%
\providecommand \url  [0]{\begingroup\@sanitize \@url }%
\providecommand \@url [1]{\endgroup\@href {#1}{\urlprefix}}%
\providecommand \urlprefix [0]{URL }%
\providecommand \Eprint[0]{\href }%
\@ifxundefined \urlstyle {%
  \providecommand \doi [1]{doi:\discretionary{}{}{}#1}%
}{%
  \providecommand \doi [0]{doi:\discretionary{}{}{}\begingroup
  \urlstyle{rm}\Url }%
}%
\providecommand \doibase [0]{http://dx.doi.org/}%
\providecommand \Doi[1]{\href{\doibase#1}}%
\providecommand \bibAnnote [3]{%
  \BibitemShut{#1}%
  \begin{quotation}\noindent
    \textsc{Key:}\ #2\\\textsc{Annotation:}\ #3%
  \end{quotation}%
}%
\providecommand \bibAnnoteFile [2]{%
  \IfFileExists{#2}{\bibAnnote {#1} {#2} {\input{#2}}}{}%
}%
\providecommand \typeout [0]{\immediate \write \m@ne }%
\providecommand \selectlanguage [0]{\@gobble}%
\providecommand \bibinfo [0]{\@secondoftwo}%
\providecommand \bibfield [0]{\@secondoftwo}%
\providecommand \translation [1]{[#1]}%
\providecommand \BibitemOpen[0]{}%
\providecommand \bibitemStop [0]{}%
\providecommand \bibitemNoStop [0]{.\EOS\space}%
\providecommand \EOS [0]{\spacefactor3000\relax}%
\providecommand \BibitemShut [1]{\csname bibitem#1\endcsname}%
%</preamble>
\bibitem{arima08}%
  \BibitemOpen
  \bibfield{author}{%
  \bibinfo {author} {\bibfnamefont{T.}~\bibnamefont{Arima}},\ }%
  \bibfield{journal}{%
  \bibinfo {journal} {J. Phys.: Condens. Matter}\ }%
  \textbf{\bibinfo {volume} {20}},\ \bibinfo {pages} {434211} (\bibinfo {year}
  {2008})%
  \bibAnnoteFile{NoStop}{arima08}%
\bibitem{brown63}%
  \BibitemOpen
  \bibfield{author}{%
  \bibinfo {author} {\bibfnamefont{W.~F.}\ \bibnamefont{Brown}}, \bibinfo
  {author} {\bibfnamefont{S.}~\bibnamefont{Shtrikman}},\ and\ \bibinfo {author}
  {\bibfnamefont{D.}~\bibnamefont{Treves}},\ }%
  \bibfield{journal}{%
  \bibinfo {journal} {J. Appl. Phys.}\ }%
  \textbf{\bibinfo {volume} {34}},\ \bibinfo {pages} {1233} (\bibinfo {year}
  {1963})%
  \bibAnnoteFile{NoStop}{brown63}%
\bibitem{krichevtsov93}%
  \BibitemOpen
  \bibfield{author}{%
  \bibinfo {author} {\bibfnamefont{B.~B.}\ \bibnamefont{Krichevtsov}}, \bibinfo
  {author} {\bibfnamefont{V.~V.}\ \bibnamefont{Pavlov}}, \bibinfo {author}
  {\bibfnamefont{R.~V.}\ \bibnamefont{Pisarev}},\ and\ \bibinfo {author}
  {\bibfnamefont{V.~N.}\ \bibnamefont{Gridnev}},\ }%
  \bibfield{journal}{%
  \bibinfo {journal} {J. Phys.: Condens. Matter}\ }%
  \textbf{\bibinfo {volume} {5}},\ \bibinfo {pages} {8233} (\bibinfo {year}
  {1993})%
  \bibAnnoteFile{NoStop}{krichevtsov93}%
\bibitem{hornreich68}%
  \BibitemOpen
  \bibfield{author}{%
  \bibinfo {author} {\bibfnamefont{R.~M.}\ \bibnamefont{Hornreich}}\ and\
  \bibinfo {author} {\bibfnamefont{S.}~\bibnamefont{Shtrikman}},\ }%
  \bibfield{journal}{%
  \bibinfo {journal} {Phys. Rev.}\ }%
  \textbf{\bibinfo {volume} {171}},\ \bibinfo {pages} {1065} (\bibinfo {year}
  {1968})%
  \bibAnnoteFile{NoStop}{hornreich68}%
\bibitem{landau}%
  \BibitemOpen
  \bibfield{author}{%
  \bibinfo {author} {\bibfnamefont{L.~D.}\ \bibnamefont{Landau}}\ and\ \bibinfo
  {author} {\bibfnamefont{E.~M.}\ \bibnamefont{Lifshitz}},\ }%
  \emph{\bibinfo {title} {Electrodynamics of Continuous Media}}\ (\bibinfo
  {publisher} {Elsevier},\ \bibinfo {year} {1984})%
  \bibAnnoteFile{NoStop}{landau}%
\bibitem{barron2004}%
  \BibitemOpen
  \bibfield{author}{%
  \bibinfo {author} {\bibfnamefont{L.~D.}\ \bibnamefont{Barron}},\ }%
  \emph{\bibinfo {title} {Molecular Light Scattering and Optical Activity}}\
  (\bibinfo {publisher} {Cambridge University Press},\ \bibinfo {address}
  {Cambridge},\ \bibinfo {year} {2004})%
  \bibAnnoteFile{NoStop}{barron2004}%
\bibitem{raab2005}%
  \BibitemOpen
  \bibfield{author}{%
  \bibinfo {author} {\bibfnamefont{R.~E.}\ \bibnamefont{Raab}}\ and\ \bibinfo
  {author} {\bibfnamefont{O.~L.}\ \bibnamefont{{De Lange}}},\ }%
  \emph{\bibinfo {title} {Multipole Theory in Electromagnetism}}\ (\bibinfo
  {publisher} {Clarendon Press},\ \bibinfo {address} {Oxford},\ \bibinfo {year}
  {2005})%
  \bibAnnoteFile{NoStop}{raab2005}%
\bibitem{natori75}%
  \BibitemOpen
  \bibfield{author}{%
  \bibinfo {author} {\bibfnamefont{K.}~\bibnamefont{Natori}},\ }%
  \bibfield{journal}{%
  \bibinfo {journal} {J. Phys. Soc. Jpn.}\ }%
  \textbf{\bibinfo {volume} {39}},\ \bibinfo {pages} {1013} (\bibinfo {year}
  {1975})%
  \bibAnnoteFile{NoStop}{natori75}%
\bibitem{zhong92}%
  \BibitemOpen
  \bibfield{author}{%
  \bibinfo {author} {\bibfnamefont{H.}~\bibnamefont{Zhong}}, \bibinfo {author}
  {\bibfnamefont{Z.~H.}\ \bibnamefont{Levine}}, \bibinfo {author}
  {\bibfnamefont{D.~C.}\ \bibnamefont{Allan}},\ and\ \bibinfo {author}
  {\bibfnamefont{J.~W.}\ \bibnamefont{Wilkins}},\ }%
  \bibfield{journal}{%
  \bibinfo {journal} {Phys. Rev. Lett.}\ }%
  \textbf{\bibinfo {volume} {69}},\ \bibinfo {pages} {379} (\bibinfo {year}
  {1992})%
  \bibAnnoteFile{NoStop}{zhong92}%
\bibitem{zhong93}%
  \BibitemOpen
  \bibfield{author}{%
  \bibinfo {author} {\bibfnamefont{H.}~\bibnamefont{Zhong}}, \bibinfo {author}
  {\bibfnamefont{Z.~H.}\ \bibnamefont{Levine}}, \bibinfo {author}
  {\bibfnamefont{D.~C.}\ \bibnamefont{Allan}},\ and\ \bibinfo {author}
  {\bibfnamefont{J.~W.}\ \bibnamefont{Wilkins}},\ }%
  \bibfield{journal}{%
  \bibinfo {journal} {Phys. Rev. B}\ }%
  \textbf{\bibinfo {volume} {48}},\ \bibinfo {pages} {1384} (\bibinfo {year}
  {1993})%
  \bibAnnoteFile{NoStop}{zhong93}%
\bibitem{muthkumar-prb96}%
  \BibitemOpen
  \bibfield{author}{%
  \bibinfo {author} {\bibfnamefont{V.~N.}\ \bibnamefont{Muthukumar}}, \bibinfo
  {author} {\bibfnamefont{R.}~\bibnamefont{Valenti}},\ and\ \bibinfo {author}
  {\bibfnamefont{C.}~\bibnamefont{Gros}},\ }%
  \bibfield{journal}{%
  \bibinfo {journal} {Phys. Rev. B}\ }%
  \textbf{\bibinfo {volume} {54}},\ \bibinfo {pages} {433} (\bibinfo {year}
  {1996})%
  \bibAnnoteFile{NoStop}{muthkumar-prb96}%
\bibitem{igarashi-prb09}%
  \BibitemOpen
  \bibfield{author}{%
  \bibinfo {author} {\bibfnamefont{J.-I.}\ \bibnamefont{Igarashi}}\ and\
  \bibinfo {author} {\bibfnamefont{T.}~\bibnamefont{Nagao}},\ }%
  \bibfield{journal}{%
  \bibinfo {journal} {Phys. Rev. B}\ }%
  \textbf{\bibinfo {volume} {80}},\ \bibinfo {pages} {054418} (\bibinfo {year}
  {2009})%
  \bibAnnoteFile{NoStop}{igarashi-prb09}%
\bibitem{blount62}%
  \BibitemOpen
  \bibfield{author}{%
  \bibinfo {author} {\bibfnamefont{E.~I.}\ \bibnamefont{Blount}},\ }%
  in\ \emph{\bibinfo {booktitle} {Solid State Physics}},\ Vol.~\bibinfo
  {volume} {13},\ \bibinfo {editor} {edited by\ \bibinfo {editor}
  {\bibfnamefont{F.}~\bibnamefont{Seitz}}\ and\ \bibinfo {editor}
  {\bibfnamefont{D.}~\bibnamefont{Turnbull}}}\ (\bibinfo {publisher}
  {Academic},\ \bibinfo {address} {New York},\ \bibinfo {year} {1962})%
  \bibAnnoteFile{NoStop}{blount62}%
\bibitem{King-Smith}%
  \BibitemOpen
  \bibfield{author}{%
  \bibinfo {author} {\bibfnamefont{R.~D.}\ \bibnamefont{King-Smith}}\ and\
  \bibinfo {author} {\bibfnamefont{D.}~\bibnamefont{Vanderbilt}},\ }%
  \bibfield{journal}{%
  \bibinfo {journal} {Phys.\ Rev.\ B}\ }%
  \textbf{\bibinfo {volume} {47}},\ \bibinfo {pages} {1651} (\bibinfo {year}
  {1993})%
  \bibAnnoteFile{NoStop}{King-Smith}%
\bibitem{resta-review07}%
  \BibitemOpen
  \bibfield{author}{%
  \bibinfo {author} {\bibfnamefont{R.}~\bibnamefont{Resta}}\ and\ \bibinfo
  {author} {\bibfnamefont{D.}~\bibnamefont{Vanderbilt}},\ }%
  in\ \emph{\bibinfo {booktitle} {Physics of Ferroelectrics: A Modern
  Perspective}},\ \bibinfo {editor} {edited by\ \bibinfo {editor}
  {\bibfnamefont{K.~M.}\ \bibnamefont{Rabe}}, \bibinfo {editor}
  {\bibfnamefont{C.~H.}\ \bibnamefont{Ahn}},\ and\ \bibinfo {editor}
  {\bibfnamefont{J.-M.}\ \bibnamefont{Triscone}}}\ (\bibinfo {publisher}
  {Springer-Verlag},\ \bibinfo {address} {Berlin},\ \bibinfo {year} {2007})\
  pp.\ \bibinfo {pages} {31--68}%
  \bibAnnoteFile{NoStop}{resta-review07}%
\bibitem{timo05}%
  \BibitemOpen
  \bibfield{author}{%
  \bibinfo {author} {\bibfnamefont{T.}~\bibnamefont{Thonhauser}}, \bibinfo
  {author} {\bibfnamefont{D.}~\bibnamefont{Ceresoli}}, \bibinfo {author}
  {\bibfnamefont{D.}~\bibnamefont{Vanderbilt}},\ and\ \bibinfo {author}
  {\bibfnamefont{R.}~\bibnamefont{Resta}},\ }%
  \bibfield{journal}{%
  \bibinfo {journal} {Phys. Rev. Lett.}\ }%
  \textbf{\bibinfo {volume} {95}},\ \bibinfo {pages} {137205} (\bibinfo {year}
  {2005})%
  \bibAnnoteFile{NoStop}{timo05}%
\bibitem{xiao05}%
  \BibitemOpen
  \bibfield{author}{%
  \bibinfo {author} {\bibfnamefont{D.}~\bibnamefont{Xiao}}, \bibinfo {author}
  {\bibfnamefont{J.}~\bibnamefont{Shi}},\ and\ \bibinfo {author}
  {\bibfnamefont{Q.}~\bibnamefont{Niu}},\ }%
  \bibfield{journal}{%
  \bibinfo {journal} {Phys. Rev. Lett.}\ }%
  \textbf{\bibinfo {volume} {95}},\ \bibinfo {pages} {137204} (\bibinfo {year}
  {2005})%
  \bibAnnoteFile{NoStop}{xiao05}%
\bibitem{ceresoli06}%
  \BibitemOpen
  \bibfield{author}{%
  \bibinfo {author} {\bibfnamefont{D.}~\bibnamefont{Ceresoli}}, \bibinfo
  {author} {\bibfnamefont{T.}~\bibnamefont{Thonhauser}}, \bibinfo {author}
  {\bibfnamefont{D.}~\bibnamefont{Vanderbilt}},\ and\ \bibinfo {author}
  {\bibfnamefont{R.}~\bibnamefont{Resta}},\ }%
  \bibfield{journal}{%
  \bibinfo {journal} {Phys. Rev. B}\ }%
  \textbf{\bibinfo {volume} {74}},\ \bibinfo {pages} {024408} (\bibinfo {year}
  {2006})%
  \bibAnnoteFile{NoStop}{ceresoli06}%
\bibitem{shi07}%
  \BibitemOpen
  \bibfield{author}{%
  \bibinfo {author} {\bibfnamefont{J.}~\bibnamefont{Shi}}, \bibinfo {author}
  {\bibfnamefont{G.}~\bibnamefont{Vignale}}, \bibinfo {author}
  {\bibfnamefont{D.}~\bibnamefont{Xiao}},\ and\ \bibinfo {author}
  {\bibfnamefont{Q.}~\bibnamefont{Niu}},\ }%
  \bibfield{journal}{%
  \bibinfo {journal} {Phys. Rev. Lett.}\ }%
  \textbf{\bibinfo {volume} {99}},\ \bibinfo {pages} {197202} (\bibinfo {year}
  {2007})%
  \bibAnnoteFile{NoStop}{shi07}%
\bibitem{malash2010}%
  \BibitemOpen
  \bibfield{author}{%
  \bibinfo {author} {\bibfnamefont{A.}~\bibnamefont{Malashevich}}, \bibinfo
  {author} {\bibfnamefont{I.}~\bibnamefont{Souza}}, \bibinfo {author}
  {\bibfnamefont{S.}~\bibnamefont{Coh}},\ and\ \bibinfo {author}
  {\bibfnamefont{D.}~\bibnamefont{Vanderbilt}},\ }%
  \bibfield{journal}{%
  \bibinfo {journal} {New J. Phys.}\ }%
  \textbf{\bibinfo {volume} {12}},\ \bibinfo {pages} {053032} (\bibinfo {year}
  {2010})%
  \bibAnnoteFile{NoStop}{malash2010}%
\bibitem{essin2010}%
  \BibitemOpen
  \bibfield{author}{%
  \bibinfo {author} {\bibfnamefont{A.~M.}\ \bibnamefont{Essin}}, \bibinfo
  {author} {\bibfnamefont{A.~M.}\ \bibnamefont{Turner}}, \bibinfo {author}
  {\bibfnamefont{J.~E.}\ \bibnamefont{Moore}},\ and\ \bibinfo {author}
  {\bibfnamefont{D.}~\bibnamefont{Vanderbilt}},\ }%
  \bibfield{journal}{%
  \bibinfo {journal} {Phys. Rev. B}\ }%
  \textbf{\bibinfo {volume} {81}},\ \bibinfo {pages} {205104} (\bibinfo {year}
  {2010})%
  \bibAnnoteFile{NoStop}{essin2010}%
\bibitem{melrose}%
  \BibitemOpen
  \bibfield{author}{%
  \bibinfo {author} {\bibfnamefont{D.~B.}\ \bibnamefont{Melrose}}\ and\
  \bibinfo {author} {\bibfnamefont{R.~C.}\ \bibnamefont{McPhedran}},\ }%
  \emph{\bibinfo {title} {Electromagnetic Processes in Dispersive Media}}\
  (\bibinfo {publisher} {Cambridge University Press},\ \bibinfo {address}
  {Cambridge},\ \bibinfo {year} {1991})%
  \bibAnnoteFile{NoStop}{melrose}%
\bibitem{explan-onsager}%
  \BibitemOpen
  \bibinfo {note} {This can be seen by expanding in $\q$ the Onsager
  reciprocity relation\cite{landau,melrose}
  $\sigma_{ab}(\q,\w;\mord)=\sigma_{ba}(-\q,\w;-\mord)$, where $\mord$ denotes
  the magnetic order parameter of the medium.}%
  \bibAnnoteFile{Stop}{explan-onsager}%
\bibitem{train08}%
  \BibitemOpen
  \bibfield{author}{%
  \bibinfo {author} {\bibfnamefont{C.}~\bibnamefont{Train}}, \bibinfo {author}
  {\bibfnamefont{R.}~\bibnamefont{Gheorghe}}, \bibinfo {author}
  {\bibfnamefont{V.}~\bibnamefont{Krstic}}, \bibinfo {author}
  {\bibfnamefont{L.-M.}\ \bibnamefont{Chamoreau}}, \bibinfo {author}
  {\bibfnamefont{N.~S.}\ \bibnamefont{Ovanesyan}}, \bibinfo {author}
  {\bibfnamefont{G.~L. J.~A.}\ \bibnamefont{Rikken}}, \bibinfo {author}
  {\bibfnamefont{M.}~\bibnamefont{Gruselle}},\ and\ \bibinfo {author}
  {\bibfnamefont{M.}~\bibnamefont{Verdaguer}},\ }%
  \bibfield{journal}{%
  \bibinfo {journal} {Nature Mater.}\ }%
  \textbf{\bibinfo {volume} {7}},\ \bibinfo {pages} {729} (\bibinfo {year}
  {2008})%
  \bibAnnoteFile{NoStop}{train08}%
\bibitem{buckingham71}%
  \BibitemOpen
  \bibfield{author}{%
  \bibinfo {author} {\bibfnamefont{A.~D.}\ \bibnamefont{Buckingham}}\ and\
  \bibinfo {author} {\bibfnamefont{M.~B.}\ \bibnamefont{Dunn}},\ }%
  \bibfield{journal}{%
  \Doi{10.1039/J19710001988}{\bibinfo {journal} {J. Chem. Soc. A}},\ \bibinfo
  {pages} {1988}}%
   (\bibinfo {year} {1971})%
  \bibAnnoteFile{NoStop}{buckingham71}%
\bibitem{harrison80}%
  \BibitemOpen
  \bibfield{author}{%
  \bibinfo {author} {\bibfnamefont{W.~A.}\ \bibnamefont{Harrison}},\ }%
  \emph{\bibinfo {title} {Solid State Theory}}\ (\bibinfo {publisher} {Dover},\
  \bibinfo {address} {New York},\ \bibinfo {year} {1980})%
  \bibAnnoteFile{NoStop}{harrison80}%
\bibitem{explan-opt-act-metals}%
  \BibitemOpen
  \bibinfo {note} {The optical activity of noncentrosymmetric metals was
  recently studied theoretically by V. P. Mineev and Yu. Yoshioka, Phys. Rev. B
  {\bf 81}, 094525 (2010).}%
  \bibAnnoteFile{Stop}{explan-opt-act-metals}%
\bibitem{wang06}%
  \BibitemOpen
  \bibfield{author}{%
  \bibinfo {author} {\bibfnamefont{X.}~\bibnamefont{Wang}}, \bibinfo {author}
  {\bibfnamefont{J.~R.}\ \bibnamefont{Yates}}, \bibinfo {author}
  {\bibfnamefont{I.}~\bibnamefont{Souza}},\ and\ \bibinfo {author}
  {\bibfnamefont{D.}~\bibnamefont{Vanderbilt}},\ }%
  \bibfield{journal}{%
  \bibinfo {journal} {Phys. Rev. B}\ }%
  \textbf{\bibinfo {volume} {74}},\ \bibinfo {pages} {195118} (\bibinfo {year}
  {2006})%
  \bibAnnoteFile{NoStop}{wang06}%
\bibitem{yates07}%
  \BibitemOpen
  \bibfield{author}{%
  \bibinfo {author} {\bibfnamefont{J.~R.}\ \bibnamefont{Yates}}, \bibinfo
  {author} {\bibfnamefont{X.}~\bibnamefont{Wang}}, \bibinfo {author}
  {\bibfnamefont{D.}~\bibnamefont{Vanderbilt}},\ and\ \bibinfo {author}
  {\bibfnamefont{I.}~\bibnamefont{Souza}},\ }%
  \bibfield{journal}{%
  \bibinfo {journal} {Phys. Rev. B}\ }%
  \textbf{\bibinfo {volume} {75}},\ \bibinfo {pages} {195121} (\bibinfo {year}
  {2007})%
  \bibAnnoteFile{NoStop}{yates07}%
\end{thebibliography}

%Merlin.mbs v4.21 2009-07-09.
%

\end{document}